%% file: main.tex
\documentclass[conference]{IEEEtran}
\IEEEoverridecommandlockouts
\usepackage{cite}
\usepackage{amsmath,amssymb,amsfonts}
\usepackage{graphicx}
\usepackage[caption=false,font=footnotesize]{subfig}
\usepackage{textcomp}
\usepackage{xcolor}
\usepackage{url}
\usepackage{hyperref}
\usepackage{placeins}
\usepackage{array}
\usepackage{algorithm}
\usepackage{algpseudocode}
\def\BibTeX{{\rm B\kern-.05em{\sc i\kern-.025em b}\kern-.08em
    T\kern-.1667em\lower.7ex\hbox{E}\kern-.125emX}}

\begin{document}

\title{Prefill/Decode-Aware Evaluation of LLM Inference on Emerging AI Accelerators
\thanks{Accepted to HPAI4S'26, co-located with IEEE IPDPS 2026.}
\thanks{\copyright~2026 IEEE. Personal use of this material is permitted. Permission from IEEE must be obtained for all other uses.}}

\author{
\IEEEauthorblockN{Shun Usami\IEEEauthorrefmark{1}}
\IEEEauthorblockA{\IEEEauthorrefmark{1}\textit{Department of Computer Science} \\
\textit{San Francisco State University}\\
San Francisco, CA, 94132 \\
susami@sfsu.edu}
\and
\IEEEauthorblockN{Venkatram Vishwanath\IEEEauthorrefmark{2}}
\IEEEauthorblockA{\IEEEauthorrefmark{2}\textit{Argonne National Laboratory}\\
Lemont, IL, 60439 \\
venkat@anl.gov}
\and
\IEEEauthorblockN{E. Wes Bethel\IEEEauthorrefmark{1}\IEEEauthorrefmark{3}}
\IEEEauthorblockA{\IEEEauthorrefmark{1}\textit{Department of Computer Science} \\
\textit{San Francisco State University}\\
San Francisco, CA, 94132 \\\IEEEauthorrefmark{3}\textit{Lawrence Berkeley National Laboratory}\\
Berkeley, CA, 94720 \\
ewbethel@sfsu.edu}
}

\maketitle

\input{00_abstract}

\input{01_introduction}
\input{02_related_work}
\input{04_implementation}
\input{03_methodology}
\input{05_results}
\input{06_discussion}
\input{07_conclusion}
\input{acknowledgments}

\bibliographystyle{IEEEtran}
\bibliography{references}
\end{document}

%% file: 00_abstract.tex
\begin{abstract}
As large language models (LLMs) are increasingly deployed in latency- and cost-sensitive settings, inference efficiency has become a central systems challenge. While GPUs dominate current deployments, a growing number of AI accelerators claim advantages for LLM inference, yet it remains unclear under which conditions such accelerators outperform GPUs in practice. Recent inference systems decompose execution into Prefill and Decode phases, which exhibit distinct computational characteristics and latency metrics, commonly captured by time to first token (TTFT) and time per output token (TPOT).

This paper presents a phase-aware evaluation of LLM inference performance across GPUs and emerging AI accelerators using a common model, Llama2-7B. By separately measuring Prefill and Decode performance, we reveal that accelerator advantages differ by phase and metric. Our results show that GPUs consistently excel in the compute-intensive Prefill phase, while GroqRack achieves significantly lower TPOT during Decode (batching not currently supported). However, GPUs regain an advantage in Decode throughput as batch size increases.

These findings demonstrate that each platform exhibits distinct phase-dependent strengths. We further analyze heterogeneous Prefill/Decode disaggregation across different accelerator platforms, identifying performance gains and the workload and network conditions under which such gains are realized.

\end{abstract}

\begin{IEEEkeywords}
Large Language Models, AI Accelerators, Performance Evaluation
\end{IEEEkeywords}

%% file: 01_introduction.tex
\section{Introduction}
\label{sec:introduction}

Large Language Models (LLMs) have become a central component of modern AI systems, with applications ranging from conversational agents to code generation and scientific assistance. As model sizes and usage continue to grow, the efficiency of inference, rather than training, has emerged as a dominant systems-level concern~\cite{chittyvenkata2023survey,zhou2024survey}. In particular, latency-sensitive and cost-sensitive inference workloads place increasing pressure on the underlying hardware and system architecture.

While GPUs dominate LLM inference due to strong linear algebra performance and mature ecosystems, emerging domain-specific accelerators claim advantages for inference workloads. However, it remains unclear under what conditions, and for which components of LLM inference, such accelerators provide tangible advantages over GPUs in practice.

Recent advances in LLM serving systems have highlighted the importance of decomposing inference into two distinct phases: Prefill and Decode~\cite{zhong2024distserve}. These phases exhibit fundamentally different computational characteristics—Prefill being compute-intensive and benefiting from parallelism, while Decode is memory-bound with sequential dependencies. This Prefill/Decode (P/D) separation has been formalized in disaggregated serving architectures and modern inference frameworks.

While Prefill/Decode disaggregation has primarily been explored in GPU-based systems, it raises a natural and important question: do different AI accelerators exhibit distinct strengths in the Prefill and Decode phases, and if so, how should such differences be measured and interpreted? Answering this question is essential for evaluating the practical role of emerging accelerators and for guiding future heterogeneous Prefill/Decode disaggregation designs, where the Prefill and Decode phases are assigned to different accelerator platforms.

Standardized benchmarks such as MLPerf Inference~\cite{reddi2020mlperf} provide a common framework for evaluating LLM inference performance. However, participation from emerging AI accelerators remains limited, and reported results typically focus on aggregate metrics such as end-to-end latency or overall throughput. For example, MLPerf Inference specifies 99th-percentile compliance thresholds on time to first token (TTFT, reflecting Prefill latency) and time per output token (TPOT, reflecting per-step Decode latency)---e.g., TTFT~$\leq$~2000\,ms and TPOT~$\leq$~200\,ms for Llama~2 70B---yet each submission reports only aggregate throughput, leaving Prefill/Decode trade-offs uncharacterized across accelerators. As a result, there is a lack of systematic, phase-aware evaluation that isolates and compares Prefill and Decode behavior across heterogeneous hardware platforms.

In this work, we present a unified experimental evaluation of Prefill and Decode performance using a common LLM model, Llama2-7B, across GPU-based platforms and emerging AI accelerators. As a representative emerging accelerator, we study GroqRack to demonstrate our methodology and reveal phase-specific performance characteristics. We evaluate Prefill and Decode performance across diverse inference scenarios to understand how accelerator strengths differ by inference phase and performance metric.

Our evaluation reveals a clear phase-dependent performance asymmetry across hardware platforms. GPUs consistently excel at Prefill through batched, compute-intensive execution, while GroqRack demonstrates substantially lower per-token latency during Decode under single-request scenarios. However, GPUs recover throughput advantages as batch size increases, illustrating a fundamental latency-throughput trade-off. These findings show that no single accelerator uniformly dominates across all phases and objectives, that Prefill and Decode must be evaluated separately for meaningful hardware comparison, and that optimal hardware selection depends critically on the system's objective function.

The contributions of this paper are threefold. First, we introduce a phase-aware evaluation methodology for LLM inference that isolates Prefill and Decode behavior under controlled workload variations. Second, we empirically characterize GPUs and emerging AI accelerators, revealing phase-dependent latency–throughput trade-offs obscured by end-to-end metrics. Finally, we present a quantitative analysis of heterogeneous Prefill/Decode disaggregation that identifies performance gains and derives break-even conditions for practical deployment.

%% file: 02_related_work.tex
\section{Background and Related Work}
\label{sec:related-work}

\subsection{Prefill/Decode Disaggregation}
\label{subsec:pd_disaggregation}

LLM inference is commonly decomposed into two phases with distinct computational characteristics: a Prefill phase that processes the entire input sequence to generate the first output token, and a Decode phase that generates subsequent tokens auto-regressively~\cite{kwon2023efficient}. The Prefill phase can be fully parallelized using matrix--matrix multiplication operators, making it compute-bound. During this process, per-token key and value vectors are produced and stored as the key--value (KV) cache for reuse in later steps. In contrast, the Decode phase is inherently sequential at the token level due to its auto-regressive nature, and its frequent accesses to the KV cache make it memory-bound.

Recent serving systems exploit this phase asymmetry through Prefill/Decode disaggregation. Zhong et al.~\cite{zhong2024distserve} formalize this approach in DistServ by separating Prefill and Decode workers that scale independently. In such designs, the KV cache generated during Prefill phase must be transferred to Decode workers to enable subsequent token generation, introducing non-negligible communication overhead. Despite this cost, DistServe demonstrates that decoupling the two phases enables more efficient resource allocation and parallelism strategies aligned with their distinct latency characteristics, leading to substantially higher goodput in GPU-based clusters.

\subsection{LLM Inference Benchmarks}
\label{subsec:llm_inference_benchmarks}

MLPerf~\cite{reddi2020mlperf} is the industry-standard benchmark suite for evaluating machine learning systems. Its inference component includes LLM workloads where TTFT and TPOT are treated as latency constraints, and systems are evaluated based on the maximum sustainable throughput achievable within predefined TTFT and TPOT bounds. In addition to performance constraints, MLPerf Inference validates output correctness against reference datasets using standard text similarity metrics (e.g., ROUGE), reflecting its goal of end-to-end system evaluation under application-facing service-level objectives.

As of the latest v5.1 datacenter results, publicly reported LLM inference submissions are largely limited to GPU-based platforms from NVIDIA and AMD, with limited coverage of emerging AI accelerators.\footnote{
See the MLPerf Inference Datacenter results page:
\url{https://mlcommons.org/benchmarks/inference-datacenter/}.} While MLPerf provides a comprehensive benchmark for deployed systems, its holistic design obscures phase-specific performance characteristics. Several recent studies have examined LLM inference performance beyond MLPerf. 

LLM-Inference-Bench~\cite{chitty2024llm} provides broader hardware coverage, including SambaNova SN40L, but reports aggregate end-to-end metrics without separating Prefill and Decode phases. While DistServe focuses on GPU-based scheduling and existing benchmarks do not systematically evaluate emerging AI accelerators under a phase-aware framework, our work addresses this gap by systematically measuring Prefill and Decode performance across GPUs and specialized AI accelerators under controlled workload variations.

\subsection{Groq Architecture Overview}
GroqRack is built around GroqCards implementing Groq’s Language Processing Unit (LPU) architecture, formerly referred to as the Tensor Streaming Processor (TSP). At the chip level, the LPU executes statically scheduled programs with a fixed instruction and memory access order, avoiding caches, dynamic scheduling, and out-of-order execution commonly used in GPUs to optimize average throughput. Instead, it relies on a large, single-level on-chip SRAM scratchpad, with model weights and intermediate states explicitly allocated and managed by the compiler~\cite{abts2020think}.

This deterministic execution model extends beyond a single chip to the rack scale. In GroqRack systems, dozens of GroqCards are interconnected via a compiler-orchestrated communication fabric, allowing both computation and inter-device communication to follow a fixed, statically determined schedule~\cite{abts2022software}.

%% file: 04_implementation.tex
\section{Model Execution Differences Across Architectures}
\label{sec:implementation}

This section details the key differences in how the same LLM model forward pass is executed on NVIDIA A100 GPUs and GroqRack accelerators, which directly impact phase-specific performance.

\subsection{NVIDIA A100 Execution Model}
\label{subsec:nvidia_implementation}

On NVIDIA A100 GPUs, we use vLLM (v0.12.0), which supports batched execution. The model forward pass accepts inputs with batch size $B > 1$ and input token length $L_{\mathrm{in}} > 1$, enabling the processing of many tokens and requests in a single invocation.

As Fig.~\ref{fig:impl_prefill_a100} shows, the Prefill phase processes all input tokens in a single parallel forward pass, while the Decode phase performs autoregressive forward passes with cached KV states until $L_{\text{out}}$ tokens are generated.

\subsection{GroqRack Execution Model}
\label{subsec:groq_implementation}

The GroqRack implementation uses the Groq SDK with the model sharded across 9 nodes (72 cards total). The compiled binary accepts only a single request and processes one token per invocation.

\begin{figure}[bt]
\centering
\subfloat[NVIDIA A100]{%
  \includegraphics[width=0.23\textwidth]{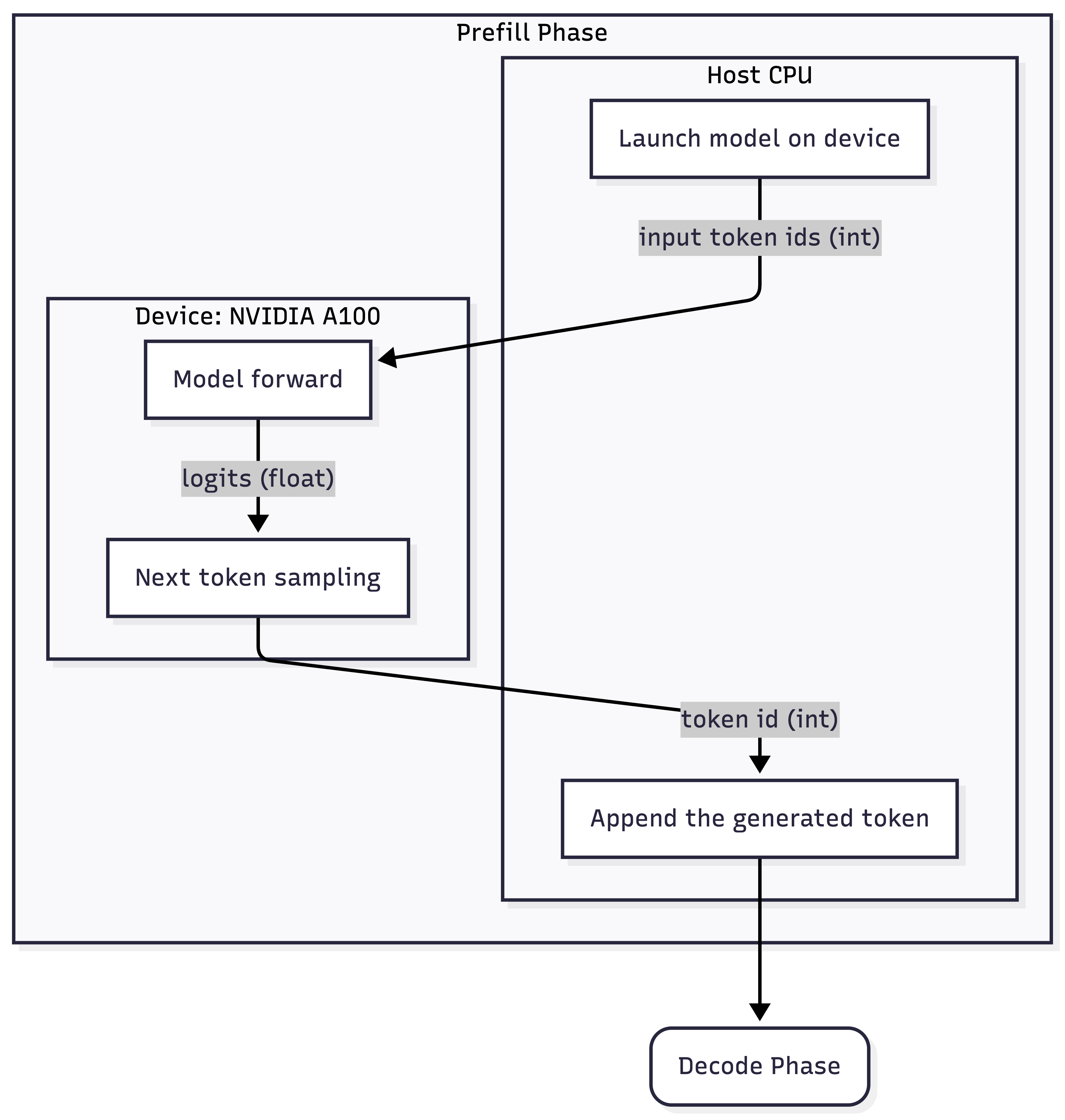}
  \label{fig:impl_prefill_a100}
}
\hfill
\subfloat[GroqRack]{%
  \includegraphics[width=0.23\textwidth]{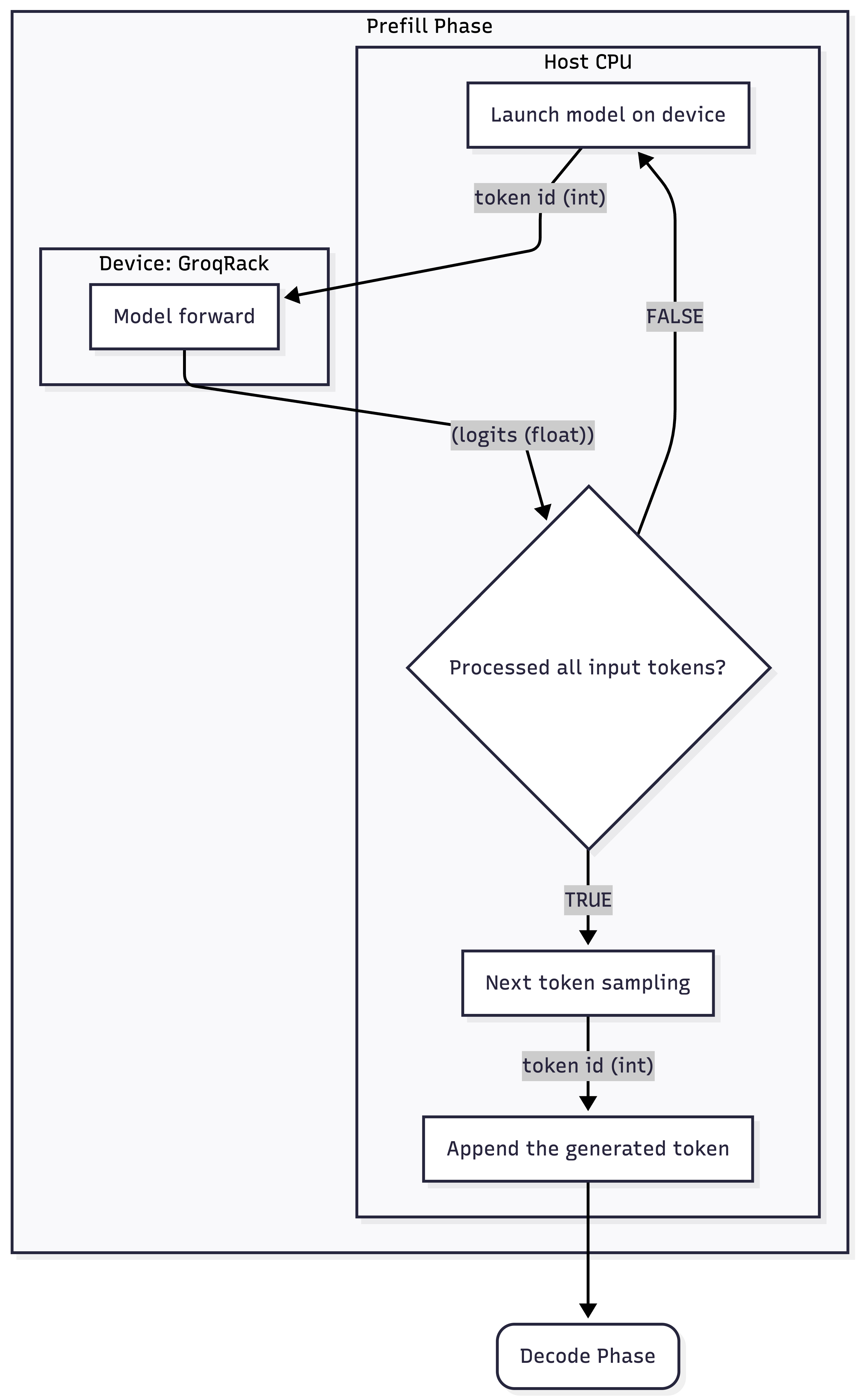}
  \label{fig:impl_prefill_groq}
}
\caption{Model execution during the Prefill phase. NVIDIA A100 processes all input tokens in a single batched forward pass, whereas the evaluated GroqRack implementation invokes the model forward pass iteratively with one token per invocation. For both platforms, the KV cache remains resident on the same device during Prefill and Decode; no cross-device KV cache transfer occurs under the evaluated configuration.}
\label{fig:impl_prefill}
\end{figure}

As Fig.~\ref{fig:impl_prefill_groq} shows, the Prefill phase processes input tokens sequentially, invoking the model forward pass once per token. After all input tokens are consumed, the first output token is sampled on the rank-0 host CPU and execution transitions to Decode. The Decode phase follows a similar iterative pattern with CPU-side sampling and MPI broadcast until $L_{\text{out}}$ tokens are generated. This sequential, single-request execution stems from GroqRack's statically scheduled, deterministic execution model with a fixed scratchpad SRAM layout. Because the LPU has no branch instructions, a compiled binary must execute with exactly the batch size it was compiled for --- it is not possible to process fewer requests than the compiled batch size, making the current single-request binary the practical choice for variable workloads.

%% file: 03_methodology.tex
\section{Methodology}
\label{sec:methodology}
Our study aims to benchmark the Prefill and Decode phases of LLM inference across heterogeneous platforms using a consistent model and workload parameters. This section details our experimental setup, including model selection, hardware platforms, inference frameworks, workload configurations, and metrics used in our experiments.

\subsection{Model Selection}
\label{subsec:model_selection}
We select Llama2-7B~\cite{touvron2023llama} as our benchmark model due to its implementation availability across multiple hardware platforms and its moderate size, which allows for efficient experimentation while still being representative of modern LLMs. The model is evaluated in its FP16 precision format, which is commonly used in inference deployments to balance performance and accuracy. With an FP16 model footprint of approximately 12.56 GiB plus KV cache requirements (512 KiB per token, up to 2 GiB for the maximum context length of 4096 tokens), the total working set of approximately 14.6 GiB fits within GroqRack's 15.42 GiB total SRAM capacity (Table~\ref{tab:hardware_specs}), making Llama2-7B nearly the largest model that can be fully accommodated on this platform.

\subsection{Hardware Platforms}
\label{subsec:hardware_platforms}

\begin{table}[bt]
\centering
\caption{Hardware Platform Specifications (based on vendor-published specifications)}
\label{tab:hardware_specs}
\renewcommand{\arraystretch}{1.4}
\begin{tabular}{p{2.0cm}|p{1.8cm}|p{3.6cm}}
\hline
\textbf{Specification} & \textbf{A100} & \textbf{GroqRack} \\
\hline
Architecture & GPU & LPU \\
\hline
System Config & Single GPU & 72 $\times$ GroqCards \\
\hline
SRAM & 40 MiB L2 & 15.42 GiB total \newline(230 MB per GroqCard) \\
\hline
DRAM & 40 GB HBM2 & --- \\
\hline
Peak Compute \newline(FP16) & 312 TFLOPS & 13.5 PFLOPS total \newline(187.5 TFLOPS per GroqCard) \\
\hline
Memory Bandwidth & 1.6 TB/s & 5.76 PB/s total \newline(80 TB/s per GroqCard) \\
\hline
\end{tabular}
\end{table}

We conduct experiments on an NVIDIA A100 SXM-40GB GPU as our baseline platform and GroqRack as a specialized AI accelerator. Table~\ref{tab:hardware_specs} summarizes the key architectural characteristics of each platform as relevant to our evaluation. For the A100 experiments, both phases are executed on the same single GPU, without hardware disaggregation. For GroqRack, both phases run on the same single GroqRack system.

\subsection{Inference Frameworks}
\label{subsec:inference_frameworks}
For NVIDIA A100, we utilize vLLM v0.12.0 due to its optimized performance for LLM inference and support for phase-aware execution. For GroqRack, we use the GroqFlow SDK and the compiled binary of Llama2-7B provided by Groq.

\subsection{Workload Configurations}
\label{subsec:workload_configurations}

We vary three workload parameters: input length ($L_{\text{in}}$), output length ($L_{\text{out}}$), and batch size ($B$). For each configuration $(L_{\text{in}}, L_{\text{out}}, B)$, all requests within a batch use prompts of equal length and generate the same number of output tokens. Input tokens are generated randomly, and EOS tokens are ignored to ensure a fixed output length.

\subsubsection{Token Lengths}
\label{subsubsec:token_length_configurations}
We evaluate $L_{\text{in}} = L_{\text{out}} \in \{100, 200, 400, 800, 1600\}$, restricting workloads to cases where input and output lengths are equal, covering short to long-context workloads while remaining within the 4096-token context limit of Llama2-7B.

\subsubsection{Batch Sizes}
\label{subsubsec:batch_size_configurations}

For NVIDIA A100, we evaluate batch sizes $B \in \{1, 2, 4, 8, 16, 32\}$ to study throughput scaling. For GroqRack, we restrict experiments to single-request execution ($B = 1$). As described in Section~\ref{sec:implementation}, GroqRack's statically scheduled execution model requires a dedicated binary per batch size. Additionally, with model weights occupying 12.56~GiB and each request requiring up to 2~GiB of KV cache, a second concurrent request would push the total to 16.56~GiB, exceeding the 15.42~GiB SRAM capacity.

\subsection{Inference Measurement Protocol}
\label{subsec:measurement_protocol}

The inference timing procedure used in this study is summarized in Algorithm~\ref{alg:measurement}. We explicitly separate the Prefill phase (first-token generation) from the autoregressive Decode phase to obtain phase-specific latency and throughput measurements.

We exclude tokenization and detokenization overhead from all timing measurements and focus solely on inference computation. Before collecting measurements, we perform warmup runs with unique input tokens to avoid KV cache reuse and to eliminate initialization overheads such as model compilation, program loading, and weight loading.

Our evaluation considers only performance metrics (latency and throughput). We do not assess output correctness or semantic quality, as the goal is to characterize phase-specific performance behavior under controlled workloads.
\begin{algorithm}[bt]
\caption{Prefill and Decode Timing Procedure}
\label{alg:measurement}
\begin{algorithmic}[1]
\State \textbf{Warmup}
\State $\_ \leftarrow \text{model}(\text{random\_inputs})$

\Statex
\State \textbf{Prefill Phase}
\State $\text{input} \leftarrow \text{GenerateRandomTokens}(L_{in})$
\State $t_{\text{prefill\_start}} \leftarrow \text{Now()}$
\State $\text{next\_tok} \leftarrow \text{model}(\text{input})$
\State $t_{\text{prefill\_end}} \leftarrow \text{Now()}$
\State $\text{input.Append}(\text{next\_tok})$

\Statex
\State \textbf{Decode Phase}
\State $t_{\text{decode\_start}} \leftarrow \text{Now()}$
\For{$i = 1$ to $L_{out}$}
  \State $\text{next\_tok} \leftarrow \text{model}(\text{input})$
  \State $\text{input.Append}(\text{next\_tok})$
\EndFor
\State $t_{\text{decode\_end}} \leftarrow \text{Now()}$
\end{algorithmic}
\end{algorithm}

\subsection{Performance Metrics}
\label{subsec:performance_metrics}

We evaluate performance using the four metrics defined in Table~\ref{tab:metrics}, where $L_{\text{in}}$, $L_{\text{out}}$, and $B$ follow the workload definitions in Section~\ref{subsec:workload_configurations}.

\begin{table}[bt]
\centering
\caption{Performance Metrics for Prefill and Decode Evaluation}
\label{tab:metrics}
\renewcommand{\arraystretch}{2.0}
\begin{tabular}{p{1.4cm}|p{2.6cm}|p{3.6cm}}
\hline
\textbf{Metric} & \textbf{Formula} & \textbf{Measures} \\
\hline
TTFT & $t_{\text{prefill\_end}} - t_{\text{prefill\_start}}$ & Latency from Prefill start to first output token generation \\
\hline
TPOT & $\dfrac{t_{\text{decode\_end}} - t_{\text{decode\_start}}}{L_{\text{out}}}$ & Average latency per generated token during Decode phase\\
\hline
Prefill Throughput & $\dfrac{L_{\text{in}} \times B}{t_{\text{prefill\_end}} - t_{\text{prefill\_start}}}$ & Input tokens processed per second during Prefill phase\\
\hline
Decode Throughput & $\dfrac{L_{\text{out}} \times B}{t_{\text{decode\_end}} - t_{\text{decode\_start}}}$ & Output tokens generated per second during Decode phase\\
\hline
\end{tabular}
\end{table}

%% file: 05_results.tex
\section{Results}
\label{sec:results}

Our experimental evaluation reveals a strong phase-dependent performance asymmetry across hardware platforms. The results demonstrate that the relative performance of accelerators depends critically on the inference phase and on whether latency or throughput is prioritized, with Prefill and Decode exhibiting fundamentally different behaviors.

\subsection{Prefill Performance}
\label{subsec:prefill_performance}

\begin{figure*}[bt]
\centering
\subfloat[TTFT]{\includegraphics[width=3.25in]{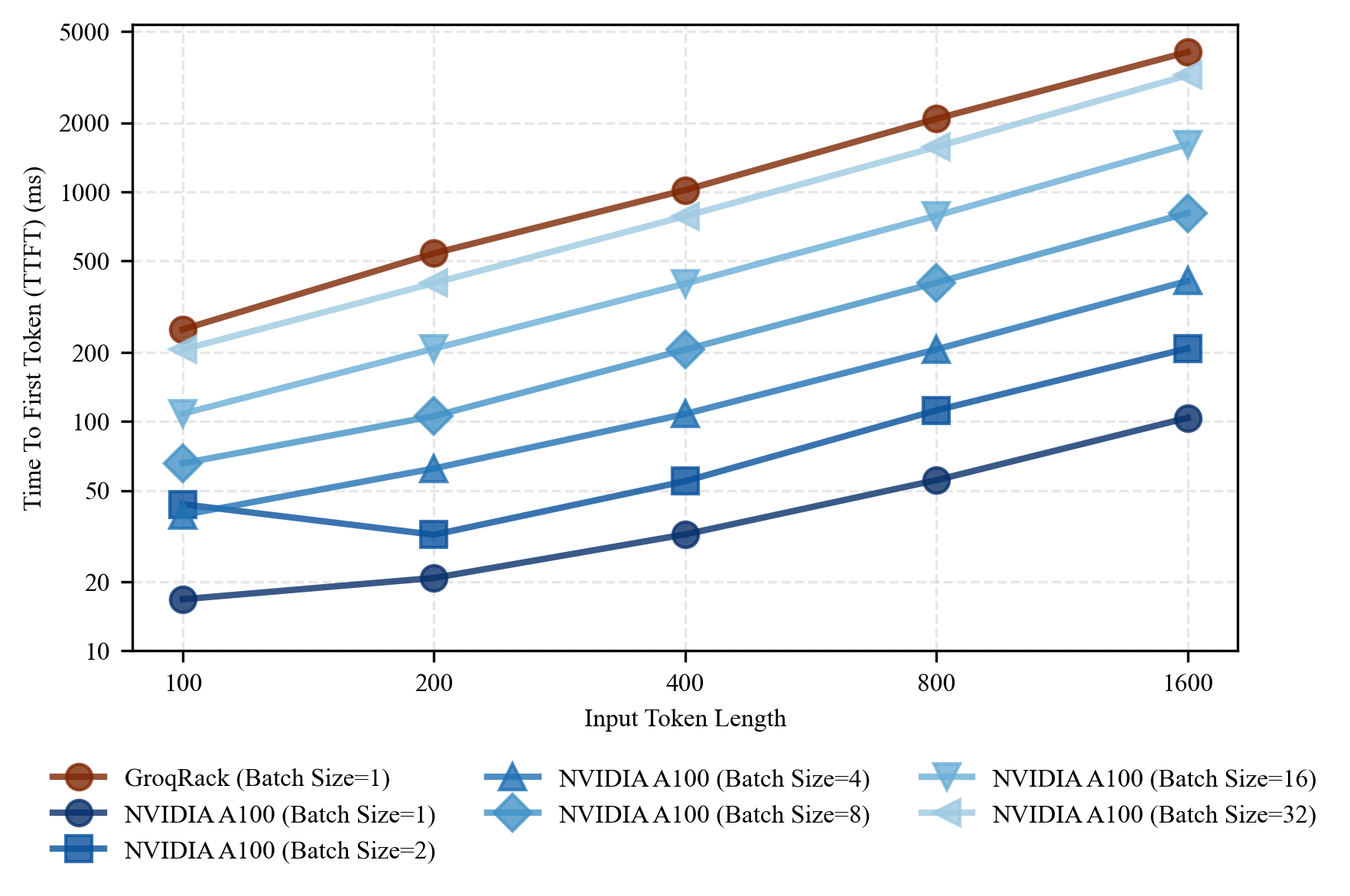}%
\label{fig:ttft}}
\hfil
\subfloat[Prefill throughput]{\includegraphics[width=3.25in]{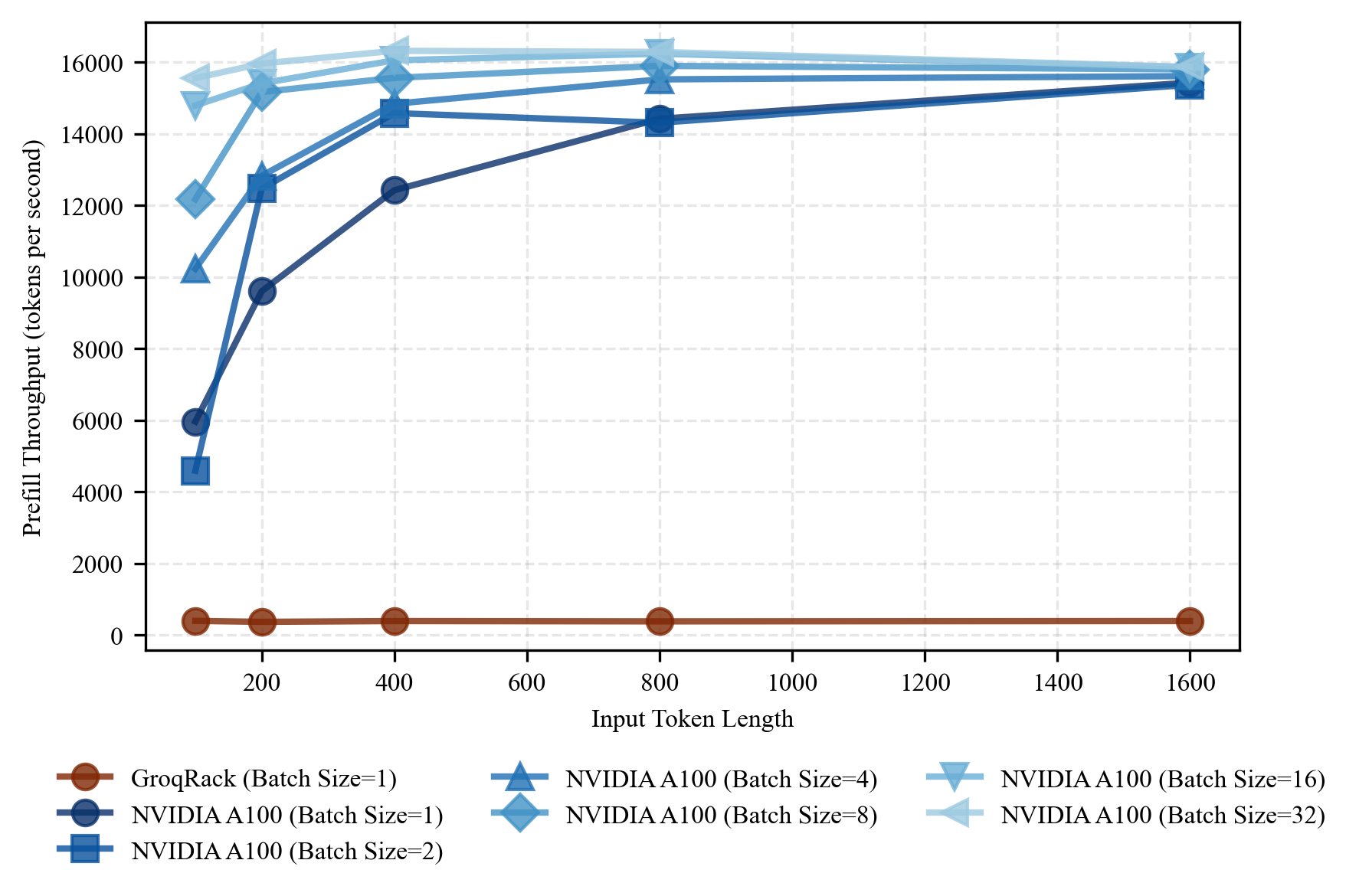}%
\label{fig:prefill_throughput}}
\caption{Prefill phase performance comparison when serving Llama2-7B on GroqRack (Batch Size 1) and NVIDIA A100 (Batch Size 1, 2, 4, 8, 16, 32).}
\label{fig:prefill}
\end{figure*}

The NVIDIA A100 GPU consistently outperforms GroqRack in the Prefill phase across all evaluated configurations. Fig.~\ref{fig:ttft} shows that the GPU achieves substantially lower TTFT than GroqRack for all input token lengths. At batch size $B=1$, GPU TTFT ranges from approximately 17~ms to 103~ms as input length increases from 100 to 1{,}600 tokens, whereas GroqRack exhibits strictly linear scaling from 252~ms to 4{,}072~ms over the same range.

As batch size increases on the GPU, TTFT grows with the total number of processed input tokens ($B \times L_{\text{in}}$). Once the GPU reaches full utilization, TTFT scales approximately linearly with workload size.

Prefill throughput results further highlight this disparity. Fig.~\ref{fig:prefill_throughput} shows that GPU throughput increases with either input token length or batch size until saturation at $B \times L_{\text{in}} \approx 800$--$1{,}600$ tokens, plateauing at approximately 15{,}000--16{,}000~tokens/s. In contrast, GroqRack maintains a nearly constant Prefill throughput of approximately 370--400~tokens/s across all evaluated input lengths.

\subsection{Decode Performance}
\label{subsec:decode_performance}

\begin{figure*}[bt]
\centering
\subfloat[TPOT]{\includegraphics[width=3.25in]{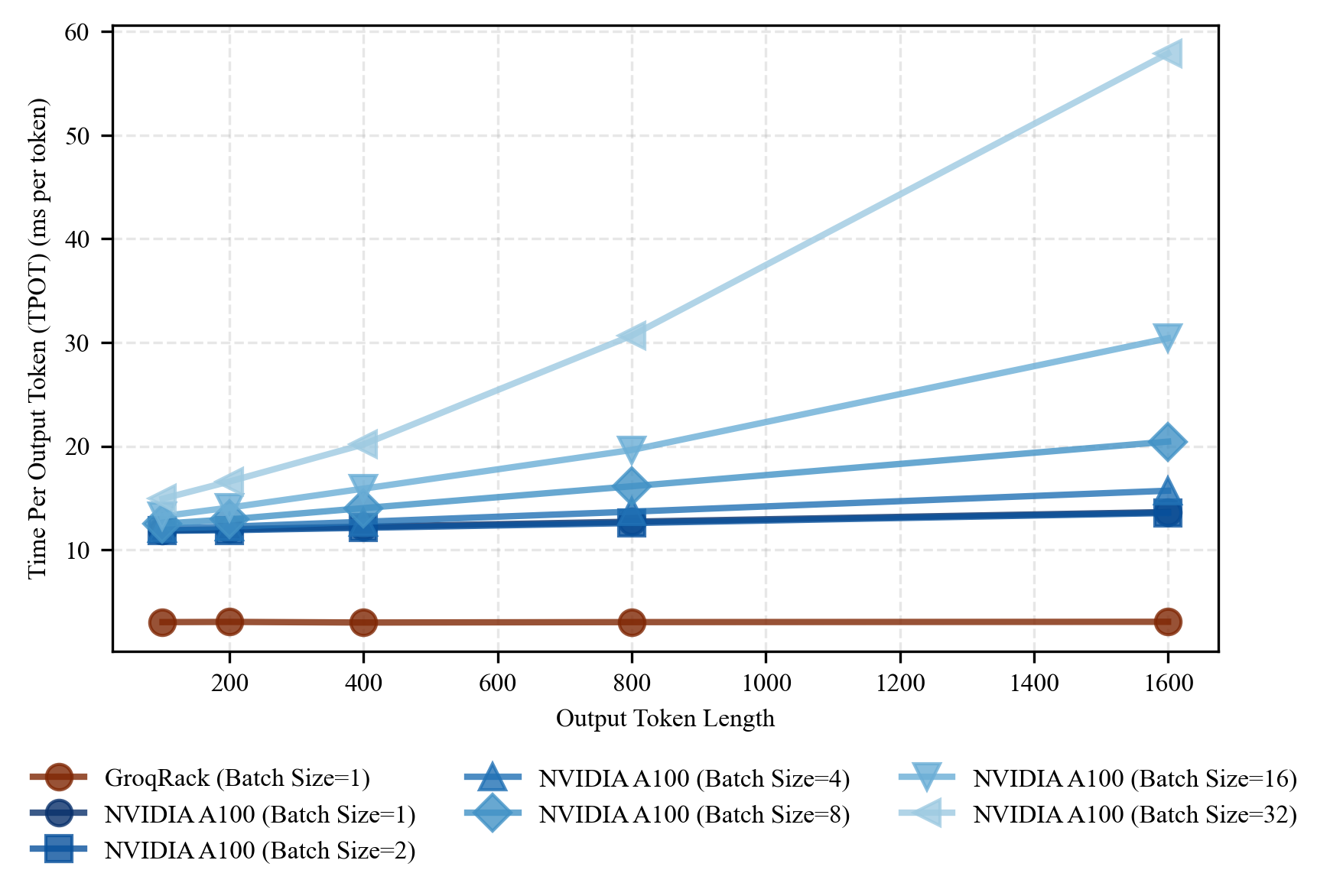}%
\label{fig:tpot}}
\hfil
\subfloat[Decode throughput]{\includegraphics[width=3.25in]{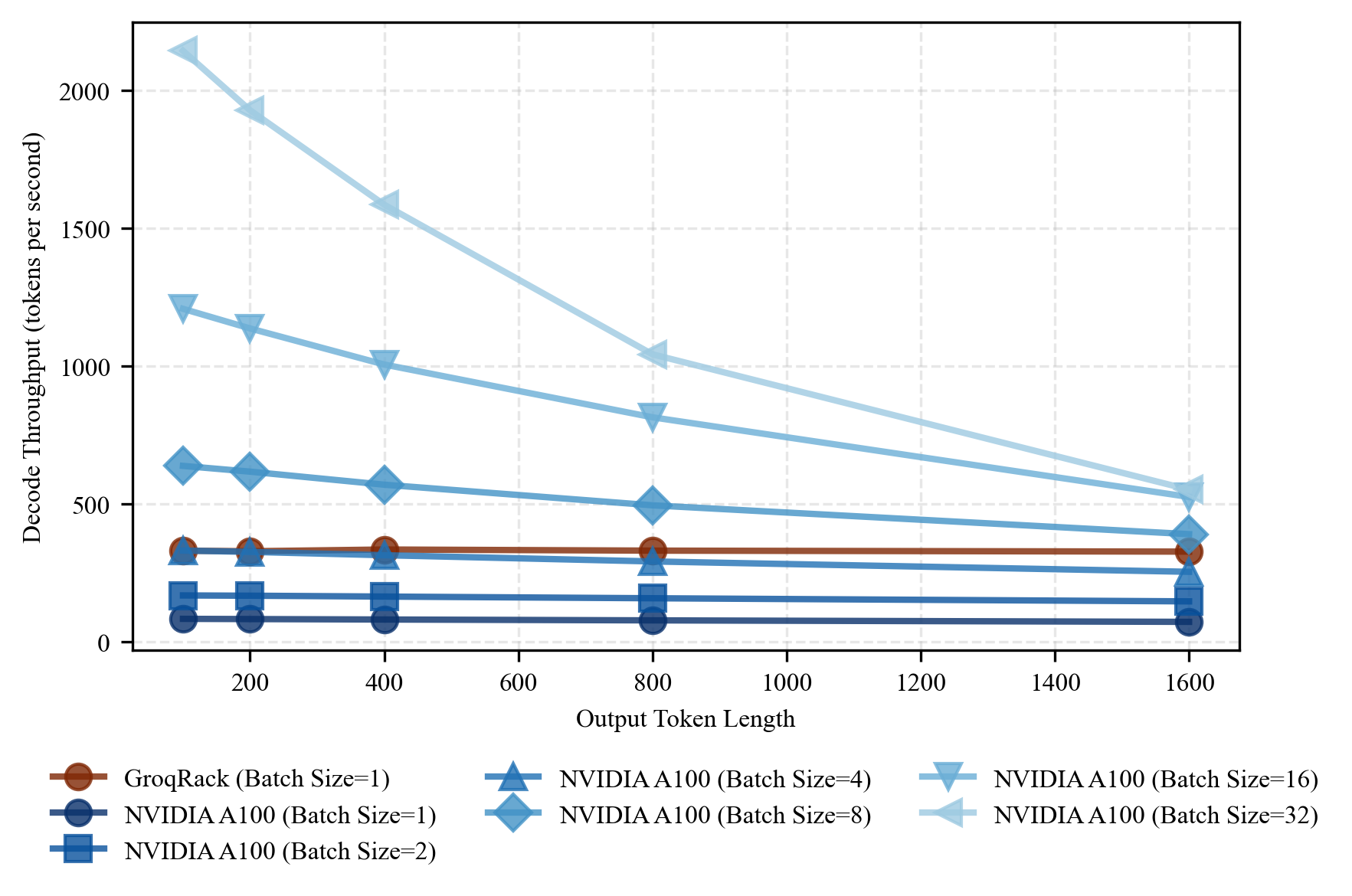}%
\label{fig:decode_throughput}}
\caption{Decode phase performance comparison when serving Llama2-7B on GroqRack (Batch Size 1) and NVIDIA A100 (Batch Size 1, 2, 4, 8, 16, 32).}
\label{fig:decode}
\end{figure*}

In contrast to the Prefill phase, the Decode phase exhibits a reversed performance trend. GroqRack consistently achieves substantially lower per-token latency than NVIDIA A100. Fig.~\ref{fig:tpot} shows that GroqRack maintains a stable TPOT of approximately 3~ms across all evaluated configurations. At $B=1$, GPU TPOT ranges from 11.88~ms to 13.64~ms and increases with output sequence length.

At higher batch sizes and longer sequences, GPU TPOT increases significantly. For example, at $B=32$ and 1{,}600 input + 1{,}600 output tokens, GPU TPOT reaches 57.90~ms at the final output token for sequences of length 3{,}200, while GroqRack maintains near-constant TPOT.

Decode throughput exhibits an opposing trend. At $B=1$, GroqRack achieves 328--336~tokens/s, compared to 73--84~tokens/s on the GPU. As batch size increases, GPU throughput scales with the number of concurrent requests and surpasses GroqRack at $B>4$. At short sequences (100 input/output tokens), the GPU reaches up to 2{,}144~tokens/s at $B=32$. However, at longer sequences, GPU throughput degrades substantially, reaching 553~tokens/s at $B=32$ for 1{,}600 input/output tokens, reducing the throughput advantage to approximately 1.7$\times$ over GroqRack.

\subsection{Performance Summary}
\label{subsec:performance_summary}
Table~\ref{tab:performance_summary} summarizes the key performance metrics observed across Prefill and Decode phases. The results highlight a clear phase-dependent performance asymmetry: GPUs dominate compute-intensive Prefill workloads, while GroqRack provides substantially lower per-token latency during Decode. GPU Decode throughput benefits from batching, but this advantage diminishes at long sequence lengths.

\begin{table}[bt]
\centering
\caption{Performance Comparison Summary for Llama2-7B Inference on NVIDIA A100 GPU and GroqRack}
\label{tab:performance_summary}
\renewcommand{\arraystretch}{1.4}
\begin{tabular}{lccc}
\hline
\textbf{Metric} & \textbf{NVIDIA A100} & \textbf{GroqRack}\\
\hline
\multicolumn{4}{l}{\textit{Prefill Phase}} \\
\hline
TTFT ($B=1, L_\text{in} = 100$) & 16.8 ms & 252 ms \\
TTFT ($B=1, L_\text{in}=1{,}600$) & 103.7 ms & 4{,}072 ms \\
Throughput (peak) & 16{,}322 tok/s & 370--397 tok/s \\
\hline
\multicolumn{4}{l}{\textit{Decode Phase}} \\
\hline
TPOT ($B=1$) & 11.88--13.64 ms & 2.98--3.05 ms \\
TPOT ($B=32$) & 14.92--57.90 ms & N/A$^{\mathrm{*}}$ \\
Throughput ($B=1$) & 73--84 tok/s & 328--336 tok/s \\
Throughput ($B=32$) & 553--2{,}144 tok/s & N/A$^{\mathrm{*}}$ \\
\hline
\multicolumn{4}{p{8.4cm}}{$^{\mathrm{*}}$GroqRack supports $B=1$ only due to current implementation constraints.} \\
\end{tabular}
\end{table}

%% file: 06_discussion.tex
\section{Discussion}
\label{sec:discussion}

This section interprets the experimental results from Section~\ref{sec:results} and examines their implications for accelerator selection and system design. We first analyze the architectural factors underlying the observed performance differences, then assess the feasibility of heterogeneous Prefill/Decode disaggregation, and finally discuss broader system implications and limitations.

\subsection{Architectural Origins of Phase-Dependent Performance}
\label{subsec:arch_origins}

We begin by examining how architectural design choices and execution models influence performance across different inference phases.

\subsubsection{Why GPUs Excel at Prefill}
\label{subsubsec:gpu_prefill}
GPUs excel at the Prefill phase because their architectures are optimized for large-scale data-parallel execution of dense linear algebra. During Prefill, all input tokens across the batch can be processed concurrently, allowing the workload to be expressed as large matrix--matrix multiplications whose effective size scales with $B \times L_{\mathrm{in}}$.

As $B \times L_{\mathrm{in}}$ increases, this parallelism enables GPUs to efficiently amortize fixed overheads and approach peak compute utilization, leading to high Prefill throughput once the device is saturated. 

\subsubsection{Why GroqRack Performs Poorly at Prefill}
\label{subsubsec:groq_prefill}

GroqRack exhibits limited Prefill performance because the current Llama2-7B implementation processes input tokens sequentially rather than in parallel. As detailed in Section~\ref{sec:implementation}, Prefill execution on GroqRack follows a token-by-token execution model similar to Decode, resulting in strictly linear TTFT scaling and nearly constant Prefill throughput.

\subsubsection{Why GroqRack Excels at Decode Latency}
\label{subsubsec:groq_decode_latency}

GroqRack achieves low and stable Decode latency because Llama2-7B inference executes entirely on explicitly managed SRAM without any cache hierarchy. The model weights (12.56~GiB) and KV cache (up to 2~GiB) together fit within the statically allocated 15.42~GiB of distributed SRAM across GroqCards (Section~\ref{subsec:model_selection}), enabling deterministic, token-by-token execution.

As a result, Decode latency is governed by fixed execution schedules rather than dynamic memory behavior, leading to near-constant TPOT across all evaluated sequence lengths.

\subsubsection{Why GPU Decode Throughput Saturates and Degrades}
\label{subsubsec:gpu_decode_throughput}
While GPUs can achieve high Decode throughput by batching requests, this advantage diminishes as batch size and sequence length increase. In this regime, each Decode step to generate the next token must access a larger portion of the accumulated KV cache, quickly exceeding on-chip cache capacity and inducing frequent L1 and L2 cache misses. Consequently, Decode performance becomes dominated by lower-bandwidth HBM accesses rather than compute throughput.

Beyond this bandwidth-driven slowdown, Decode performance further degrades when the KV cache footprint exceeds the available GPU memory budget. At $B=32$ with 1{,}600 input and 1{,}600 output tokens, the KV cache footprint (102{,}400 tokens) exceeds the available HBM allocation (45{,}584 tokens reported by vLLM) by more than $2\times$. In this case, vLLM limits effective Decode concurrency to respect KV cache capacity, leading to run-time stalling and request queueing. This working set size constraint explains the observed Decode throughput saturation and degradation in Section~\ref{subsec:decode_performance}.

\subsection{Heterogeneous Prefill/Decode Disaggregation Analysis}
\label{subsec:heterogeneous_analysis}

While heterogeneous Prefill/Decode disaggregation is conceptually appealing, its practical viability depends on the balance between performance gains and system-level overheads, particularly KV cache transfer costs.

\subsubsection{KV Cache Transfer Overhead}
\label{subsubsec:kv_cache_transfer}
KV cache transfer latency per input token is modeled as:
\begin{equation}
t_{\text{transfer}} =
\frac{\text{KV cache size per input token}}
     {\text{network bandwidth} \times \text{bandwidth efficiency}}.\label{eq:kv_overhead}
\end{equation}
For Llama2-7B, the KV cache size per input token is 512~KiB, and we assume a bandwidth efficiency of 0.8 throughout this analysis.

We consider three representative network configurations: a 25~Gbps Ethernet baseline ($t_{\text{transfer}} = 0.210$~ms/token); a 100~Gbps-per-GPU setting reflecting realistic Slingshot-based deployments ($t_{\text{transfer}} = 0.052$~ms/token); and a 1.8~TB/s NVLink configuration, reflecting the bandwidth scale of next-generation GPU interconnects, as an optimistic upper bound ($t_{\text{transfer}} = 0.00036$~ms/token).\footnote{ The 1.8~TB/s NVLink bandwidth corresponds to the per-direction specification of next-generation NVIDIA NVLink (6th generation), as announced for the Rubin architecture. While NVLink is not applicable to inter-rack GPU--GroqRack communication, it is used here as a theoretical lower bound on KV cache transfer overhead under extremely high-bandwidth interconnect assumptions. }

\subsubsection{End-to-end Latency Analysis}
\label{subsubsec:e2e_latency}
\begin{figure}[bt]
\centering
\includegraphics[width=\columnwidth]{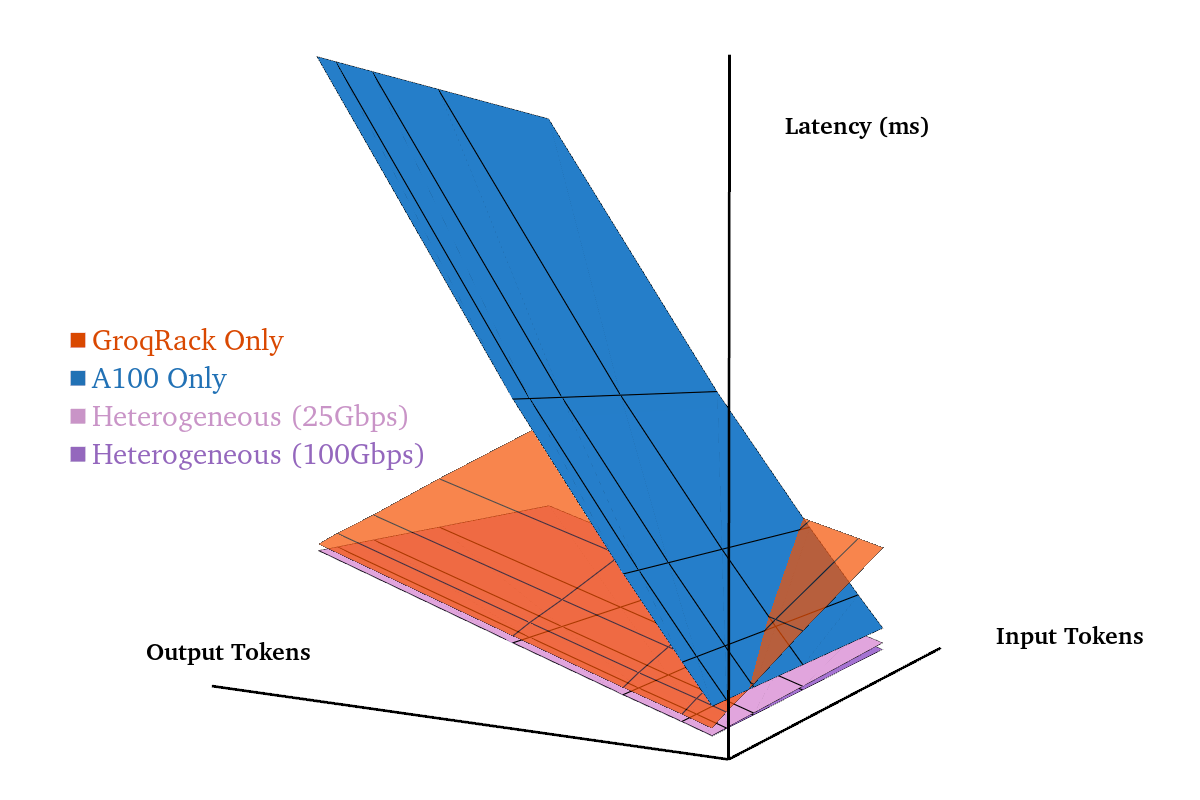}
\caption{End-to-end latency comparison for Llama2-7B across homogeneous (A100-only, GroqRack-only) and heterogeneous (Prefill on A100, Decode on GroqRack) inference strategies with two network configurations: 25~Gbps Ethernet and 100~Gbps HPE Slingshot.}
\label{fig:hetero_arch_comparison}
\end{figure}

Fig.~\ref{fig:hetero_arch_comparison} shows a model-based end-to-end latency comparison between single-platform inference and heterogeneous Prefill/Decode disaggregation. In the workload range shown, heterogeneous disaggregation (Prefill on A100, Decode on GroqRack) consistently achieves the lowest end-to-end latency for both 25~Gbps Ethernet and 100~Gbps HPE Slingshot. This result reflects the combination of GPU-efficient Prefill and GroqRack's low per-token Decode latency, with KV cache transfer contributing a secondary effect within this regime.

\begin{figure}[bt]
\centering
\subfloat[$L_{\mathrm{in}} = 100$, $L_{\mathrm{out}} = 1$]{\includegraphics[width=0.48\linewidth]{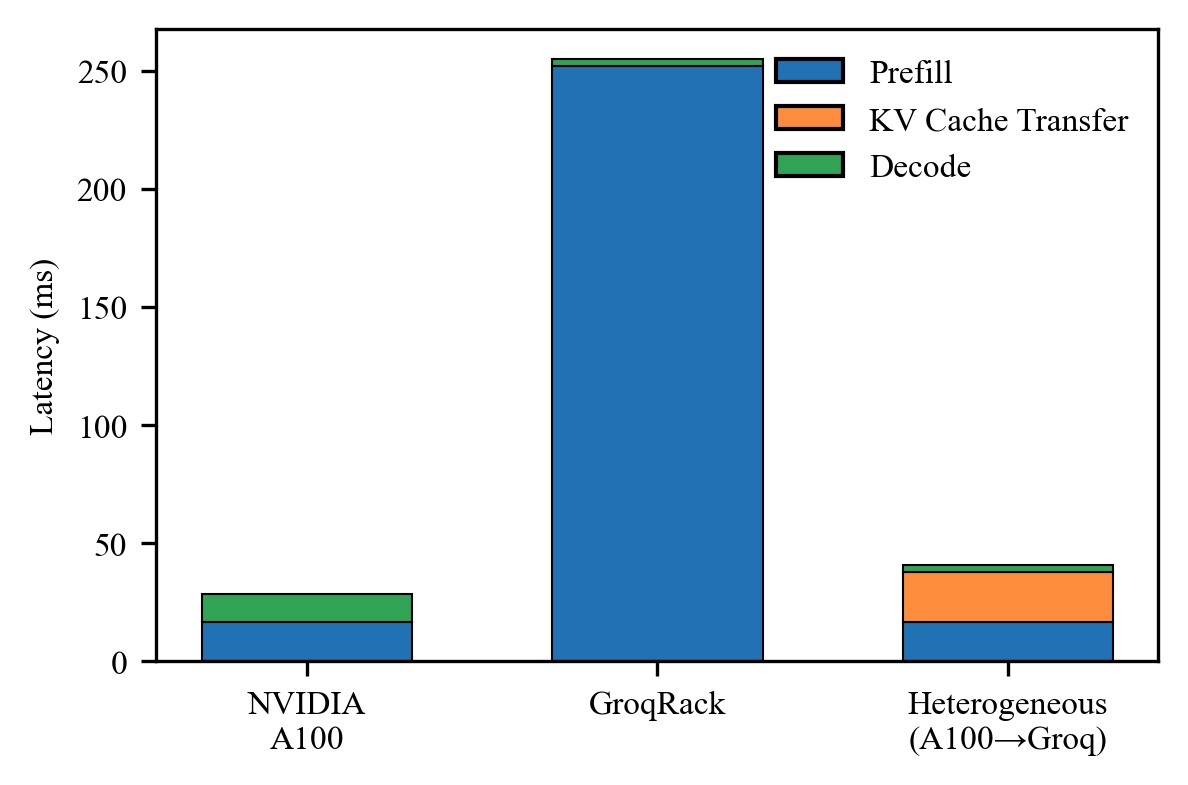}%
\label{fig:hetero_stack_100_1}}
\hfill
\subfloat[$L_{\mathrm{in}} = 100$, $L_{\mathrm{out}} = 100$]{\includegraphics[width=0.48\linewidth]{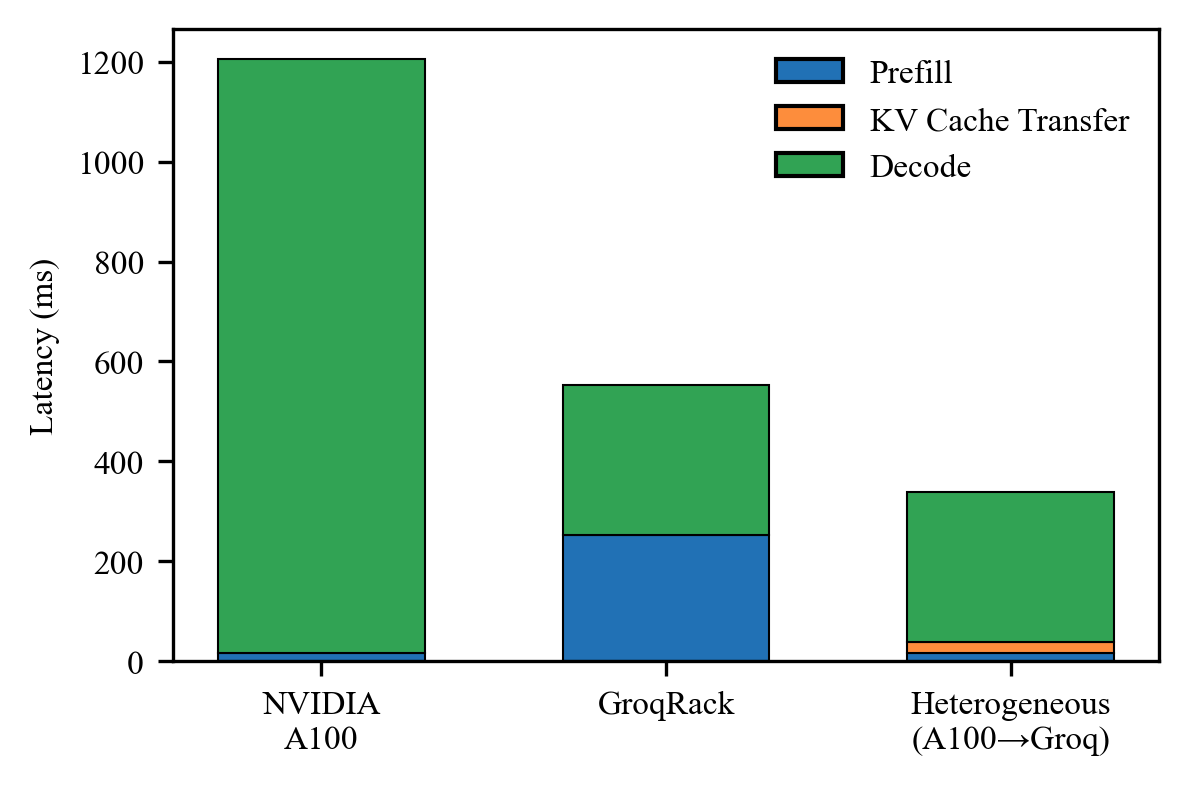}%
\label{fig:hetero_stack_100_100}}
\caption{Latency breakdown comparison across homogeneous (A100-only, GroqRack-only) and heterogeneous (Prefill on A100, Decode on GroqRack) architectures illustrating workload-dependent performance. KV cache transfer assumes 25~Gbps Ethernet.}
\label{fig:hetero_stack_comparison}
\end{figure}

To complement this aggregate view, Fig.~\ref{fig:hetero_stack_comparison} presents latency breakdowns for representative workloads with different input--output length ratios.
Fig.~\ref{fig:hetero_stack_100_1} corresponds to an extreme prefill-heavy workload ($L_{\mathrm{in}}=100$, $L_{\mathrm{out}}=1$), where the Decode phase is too short to amortize the KV cache transfer cost. In this regime, the KV cache transfer overhead outweighs the Decode latency reduction achieved by GroqRack, rendering heterogeneous disaggregation unfavorable.
In contrast, Fig.~\ref{fig:hetero_stack_100_100} shows that for balanced workloads ($L_{\mathrm{in}}=100$, $L_{\mathrm{out}}=100$), the transfer overhead constitutes only a small fraction of end-to-end latency, and heterogeneous disaggregation consistently outperforms homogeneous baselines.

\subsubsection{Heterogeneous Disaggregation Always Outperforms GroqRack-Only Inference}
\label{subsubsec:hetero_vs_groq}
Although both Prefill latency and KV cache transfer overhead scale linearly with $L_{\mathrm{in}}$, 
the absolute per-token Prefill latency reduction from offloading Prefill to A100 
always exceeds the per-token transfer overhead (prefill gain $\ge$ 2~ms/token, transfer overhead $\le$ 0.21~ms/token). 
Consequently, heterogeneous disaggregation remains strictly favorable for all input lengths, and no break-even point exists for this comparison.

\subsubsection{Break-Even Analysis Compared to A100-Only Inference}
\label{subsubsec:break_even_analysis}
A non-trivial trade-off arises when comparing heterogeneous disaggregation against A100-only inference. While heterogeneous disaggregation reduces Decode latency by offloading Decode to GroqRack, it incurs additional KV cache transfer overhead. Heterogeneous disaggregation is beneficial when the Decode latency gain amortizes the transfer cost, yielding
\begin{equation}
L_{\mathrm{in}} \times t_{\text{transfer}} < L_{\mathrm{out}} \times t_{\text{decode\_gain}}.\label{eq:break_even}
\end{equation}

Using experimentally measured Prefill and Decode latencies together with network-dependent transfer cost estimates, we derive break-even thresholds summarized in Table~\ref{tab:break_even}.

Heterogeneous Prefill/Decode disaggregation is advantageous for the majority of practical workloads, even when using a 25~Gbps Ethernet link. The only unfavorable regime corresponds to extremely input-heavy workloads, where the input length exceeds the output length by roughly 50$\times$ or more. Such cases typically arise in tasks that produce minimal outputs, such as binary decisions or categorical classification based on large input contexts. For most other workloads, heterogeneous disaggregation is preferable. Moreover, with extremely high-bandwidth interconnects such as NVLink (used here only as a theoretical upper bound rather than a feasible GPU–GroqRack deployment scenario), the KV cache transfer overhead becomes negligible for virtually all realistic inference scenarios.

\begin{table}[tb]
\centering
\caption{Required output--input length ratio for heterogeneous Prefill/Decode disaggregation}
\label{tab:break_even}
\renewcommand{\arraystretch}{2.0}
\begin{tabular}{p{2.4cm}|p{5.6cm}}
\hline
\textbf{Interconnect} & Heterogeneous wins when \\
\hline
25~Gbps Ethernet &
$\displaystyle \frac{L_{\mathrm{out}}}{L_{\mathrm{in}}} \;>\; [\,\frac{1}{52},\;\frac{1}{42}\,]^{\mathrm{*}}$ \\
\hline
100~Gbps Slingshot &
$\displaystyle \frac{L_{\mathrm{out}}}{L_{\mathrm{in}}} \;>\; [\,\frac{1}{203},\;\frac{1}{169}\,]^{\mathrm{*}}$ \\
\hline
NVLink (1.8~TB/s)$^{\dagger}$ &
$\displaystyle \frac{L_{\mathrm{out}}}{L_{\mathrm{in}}} \;>\; [\,\frac{1}{29{,}300},\;\frac{1}{24{,}400}\,]^{\mathrm{*}}$ \\
\hline
\multicolumn{2}{l}{$^{\mathrm{*}}$Ranges reflect variability in the measured per-token Decode gain.} \\
\multicolumn{2}{l}{$^{\dagger}$NVLink is included as a theoretical upper bound and does not represent a feasible GPU--GroqRack interconnect.} \\
\end{tabular}
\end{table}

\subsection{Implications for System Design}
\label{subsec:system_design_implications}
The phase-dependent performance asymmetry observed in our experiments has important implications for the design of practical LLM inference systems.

\subsubsection{Accelerator Specialization versus Versatility}
\label{subsubsec:latency_throughput_tradeoffs}

While GroqRack excels at Decode but performs poorly at Prefill, GPUs can efficiently handle both phases. This versatility enables GPUs to serve as overflow capacity for Decode when GroqRack systems reach saturation, preventing request queuing and maintaining system throughput under high load.

\subsubsection{Multi-Factor Routing Policies}
\label{subsubsec:workload_aware_routing}

In practice, the break-even condition in \eqref{eq:break_even} directly informs routing decisions in heterogeneous systems. Given workload characteristics $(L_{\mathrm{in}}, L_{\mathrm{out}})$ and current system conditions (accelerator utilization, network congestion), requests can be dynamically routed to heterogeneous disaggregation or single-platform execution depending on whether the expected Decode latency reduction amortizes the KV cache transfer overhead.

\subsubsection{Challenges in Efficient KV Cache Transfer}
\label{subsubsec:system_challenges}
Although KV cache transfer overhead is often amortizable in theory, efficiently enabling low-latency KV cache exchange across heterogeneous accelerators remains an open system challenge. In contrast to homogeneous GPU deployments that can leverage high-bandwidth interconnects (e.g., NVLink) and emerging point-to-point communication libraries such as NIXL, no standardized mechanism currently exists for efficient KV cache transfer between GPUs and non-GPU accelerators.

\subsection{Generalizability and Limitations}
\label{subsec:limitations}

Our evaluation targets Llama2-7B in FP16 under controlled synthetic workloads. While absolute latency values are model-specific, the qualitative trade-offs persist for transformer-based models: Prefill remains compute-bound and favors parallelism; Decode remains memory-bound and benefits from low-latency memory access. Per-token latency and KV cache size scale linearly with model dimension, preserving the structural trade-off between Decode gains and transfer costs, though quantitative break-even points shift with model size.

Our analysis assumes idealized KV cache transfer based on sustained bandwidth. Production systems face additional overheads from congestion, protocols, and multi-tenancy. We do not consider economic factors. Despite these limitations, the core insight remains: phase-dependent performance asymmetry necessitates separate evaluation of Prefill and Decode when comparing heterogeneous platforms.

Despite these limitations, the central insights remain robust: Prefill and Decode exhibit fundamentally different computational characteristics; accelerators demonstrate phase-dependent performance asymmetry; and the viability of heterogeneous disaggregation depends on balancing phase-specific gains against transfer overheads. These principles provide a general framework for evaluating heterogeneous LLM inference systems beyond the specific configuration studied here.

%% file: 07_conclusion.tex
\section{Conclusion}
\label{sec:conclusion}
This paper presents a phase-aware evaluation methodology for LLM inference that isolates Prefill and Decode performance across GPUs and emerging AI accelerators. Using Llama2-7B as a benchmark model, our systematic evaluation reveals fundamental phase-dependent performance asymmetry: GPUs achieve up to 39$\times$ lower latency and 41$\times$ higher throughput in the compute-intensive Prefill phase through massive parallelism, while GroqRack demonstrates 4$\times$ lower per-token latency in the memory-bound Decode phase through deterministic execution and large on-chip SRAM. However, GPUs recover throughput advantages as batch size increases due to effective parallel processing across independent requests. These findings demonstrate that the platforms exhibit complementary strengths across inference phases and system objectives, that aggregate end-to-end metrics can obscure critical phase-specific trade-offs, and that optimal accelerator selection depends on workload characteristics as well as whether latency or throughput is prioritized.

Building on these observations, our heterogeneous Prefill/Decode disaggregation analysis suggests that combining GPU-based Prefill with accelerator-based Decode can achieve lower end-to-end latency than homogeneous approaches when output length is sufficient to amortize KV cache transfer costs. Although this analysis is based on analytical and hypothetical estimates rather than a fully integrated system, the results indicate that heterogeneous Prefill/Decode disaggregation is a structurally promising design point rather than a marginal optimization.

Realizing such heterogeneous inference architectures in practice will require advances beyond individual accelerators. In particular, efficient and predictable GPU--accelerator communication becomes a key enabler, encompassing high-bandwidth physical interconnects as well as software stacks that expose phase boundaries, manage KV cache placement and transfer, and coordinate workload-aware scheduling across devices. As LLM workloads continue to scale in both model size and context length, we believe that co-designing hardware links, runtime systems, and orchestration layers for cross-device execution will be increasingly important for enabling efficient, scalable, and cost-effective LLM inference.

%% file: acknowledgments.tex
\section*{Acknowledgment}
This research used resources of the Argonne Leadership Computing Facility, which is a U.S. Department of Energy Office of Science User Facility operated under contract DE-AC02-06CH11357.